\documentclass[amsmath,amssymb,aps,prl,onecolumn,superscriptaddress,showpacs,floatfix,notitlepage]{revtex4-1}

\usepackage{graphicx}
\usepackage{color}
\usepackage{epstopdf}
\usepackage{braket}
\usepackage{lineno}
\usepackage{ulem}

\begin{document} 

\title{Vibron-assisted spin excitation in a magnetically anisotropic nickelocene complex}


\author{N. Bachellier}
\author{B. Verlhac}
\email{verlhac@ipcms.unistra.fr}
\affiliation{Universit\'{e} de Strasbourg, CNRS, IPCMS, UMR 7504, F-67000 Strasbourg, France}
\author{L. Garnier}
\affiliation{Universit\'{e} de Strasbourg, CNRS, IPCMS, UMR 7504, F-67000 Strasbourg, France}
\author{J. Zald\'{i}var}
\author{C. Rubio-Verd\'{u}}
\affiliation{CIC nanoGUNE, 20018 Donostia-San Sebasti\'{a}n, Spain}
\author{P. Abufager}
\affiliation{Instituto de F\'{i}sica de Rosario, Consejo Nacional de Investigaciones Cient\'{i}ficas y T\'ecnicas (CONICET) and Universidad Nacional de Rosario, Av. Pellegrini 250 (2000) Rosario, Argentina}
\author{M. Ormaza}
\affiliation{Universit\'{e} de Strasbourg, CNRS, IPCMS, UMR 7504, F-67000 Strasbourg, France}
\author{D.-J. Choi}
\affiliation{Centro de F{\'{\i}}sica de Materiales (CFM), 20018 Donostia-San Sebasti\'an, Spain}
\affiliation{Ikerbasque, Basque Foundation for Science, Bilbao, Spain}
\author{M.-L. Bocquet}
\affiliation{PASTEUR, D\'epartement de Chimie, Ecole Normale Sup\'erieure, PSL Research University, Sorbonne Universit\'es, UPMC Univ. Paris 06, CNRS, 75005 Paris, France}
\author{J.I. Pascual}
\affiliation{CIC nanoGUNE, 20018 Donostia-San Sebasti\'{a}n, Spain}
\affiliation{Ikerbasque, Basque Foundation for Science, Bilbao, Spain}
\author{N. Lorente}
\affiliation{Centro de F{\'{\i}}sica de Materiales (CFM), 20018 Donostia-San Sebasti\'an, Spain}
\affiliation{Donostia International Physics Center (DIPC), 20018 Donostia-San Sebasti\'an, Spain}
\author{L. Limot}
\email{limot@ipcms.unistra.fr}
\affiliation{Universit\'{e} de Strasbourg, CNRS, IPCMS, UMR 7504, F-67000 Strasbourg, France}
\date{\today}

\begin{abstract}
The ability to electrically-drive spin excitations in molecules with magnetic anisotropy is key for high-density storage and quantum-information technology. Electrons, however, also tunnel via the vibrational excitations unique to a molecule. The interplay of spin and vibrational excitations offers novel routes to study and, ultimately, electrically manipulate molecular magnetism. Here we use a scanning tunneling microscope to electrically induce spin and vibrational excitations in a single molecule consisting of a nickelocene magnetically coupled to a Ni atom. We evidence a vibron-assisted spin excitation at an energy one order of magnitude higher compared to the usual spin excitations of nickelocene and explain it using first-principles calculations that include electron correlations. Furthermore, we observe that spin excitations can be quenched by modifying the Ni-nickelocene coupling. Our study suggests that nickelocene-based complexes constitute a model playground for exploring the interaction of spin and vibrations in the electron transport through single magnetic molecules.
 \end{abstract}

\maketitle 

Magnetic molecules are potential candidates for information-storing technology~\cite{Mannini2010}, molecular spintronics~\cite{Bogani2008} and quantum computing~\cite{Loss2001}, provided that their axial magnetic anisotropy $DS_z^2$ ensures a magnetic bistability among long-lived magnetic states. But molecules also vibrate. In particular, in magnetic molecules the interplay of vibrational modes, or vibrons, with the spin degrees of freedom is known to impact the spin lifetime~\cite{Gatteschi2006,Ganzhorn2013,Droghetti2015}. Since vibrational modes can couple to the electric charge, producing vibron-assisted electron excitations in transport~\cite{Stipe1998,Park2000,LeRoy2004}, expectations are that similar effects should be also observed with the electronic spin~\cite{McCaskey2015,Kenawy2017}.

Concerning this last point, experimental work has predominantly focused on a well-known spin-related many-body effect, the Kondo effect. The electron-vibron interaction in Kondo molecules was shown to produce satellite Kondo resonances in the differential conductance spectra at the bias of the vibron's excitation energy~\cite{Yu2004,Fernandez-Torrente2008,Rakhmilevitch2014}. These resonances were ascribed to tunneling electrons that have their spin flipped when elastically scattering off the molecular spin, but with sufficient energy to activate a vibrational mode in the molecule~\cite{Parks2007}. The question arises whether a similar mechanism is also possible with other spin scattering mechanisms that magnetically excite a molecule such as inelastic scattering involving magnetic anisotropy~\cite{Hirjibehedin2007}. These so-called spin excitations show great potential in view of an all-electrical manipulation of the molecular spin~\cite{Loth2010,Heinrich2013}.

\begin{figure}
  \includegraphics[width=0.55\textwidth]{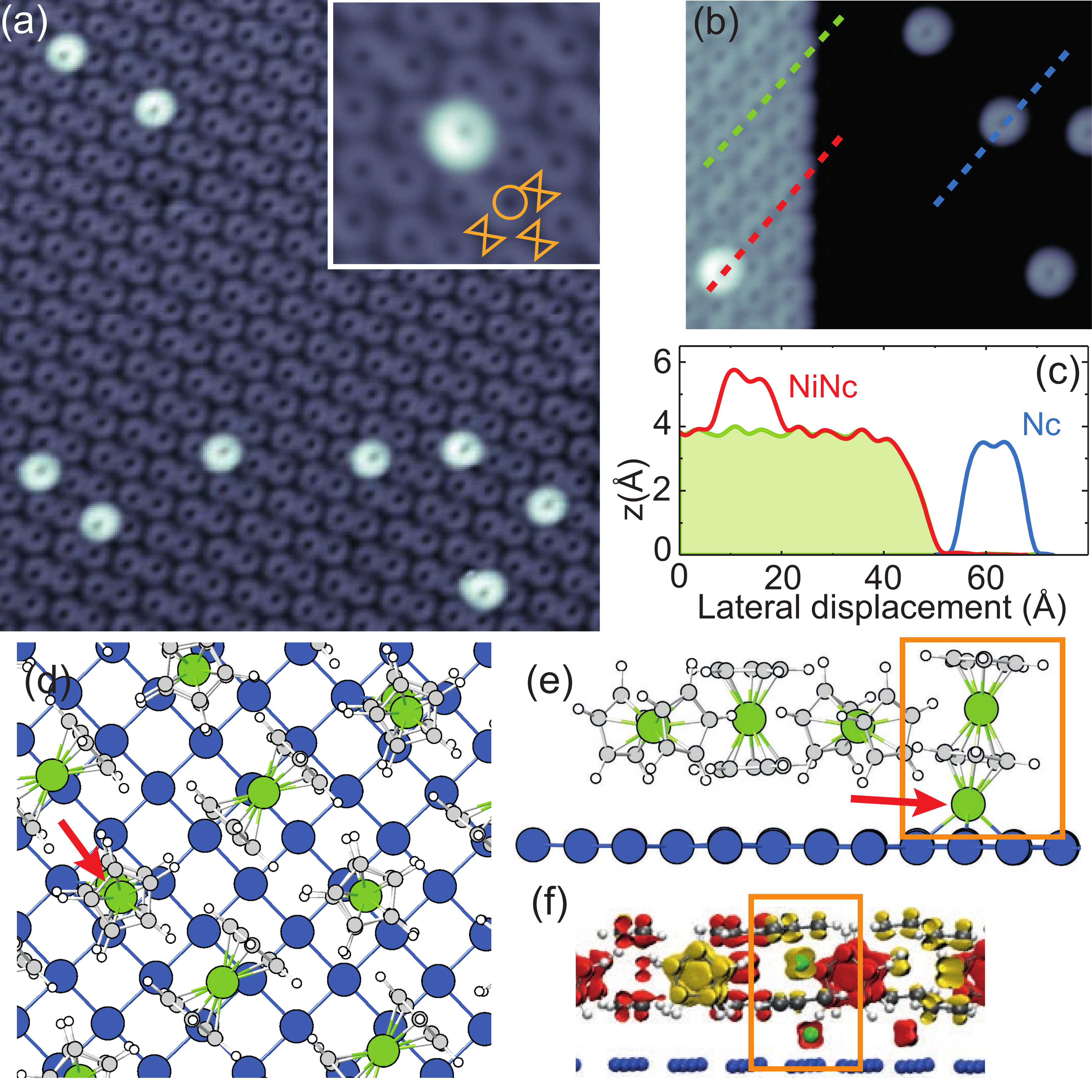}
  \caption{(a) Self-assembled layer (paired configuration) and NiNc complex on Cu(100) (image size: $16\times 16$~nm$^2$, sample bias: $20$~mV, tunneling current: $20$~pA). Inset: Close up-view of a NiNc complex in the paired layer. The molecular arrangement is sketched in orange as a circle for the vertical Nc and as a hourglass for the horizontal Nc ($4\times 4$~nm$^2$, $20$~mV, $100$~pA). (b) Edge of a paired layer and isolated Nc molecules on Cu(100) ($10\times 10$~nm$^2$, $20$~mV, $100$~pA) and (c) their corresponding height profiles along the dashed lines in (b). (d) Top view, (e) Side view of the DFT optimized structure of the NiNc  complex embedded in a Nc paired layer \textemdash some molecules have been removed for clarity (H: white, C: grey, Ni: green). Arrow indicates the Ni adatom. (f) Side view of the spin density map of the paired structure containing two Ni adatoms. Yellow rectangles mark the NiNc complex. Yellow: Spin up, red: Spin down. 
\label{fig1}}
\end{figure}

Here, we use scanning tunneling microscopy (STM) to demonstrate the presence of a combined vibrational-spin excitation in a single nickelocene molecule [Ni(C$_5$H$_5$)$_2$, see Fig.~\ref{fig1}(d); noted Nc hereafter] coupled to a Ni atom. Nickelocene is a spin $S=1$ molecule of the metallocene family with magnetic anisotropy, where spin excitations produce a sizable increase of the electronic transport~\cite{Ormaza2017a}. We show that the on-surface assembly of a nickel-nickelocene complex (NiNc, hereafter) on Cu(100) can promote a sizable vibrational mode. A second spin excitation appears then at energies comparable to those of transition atoms on thin insulating films~\cite{Rau2014,Baumann2015}. With the help of density functional theory (DFT) calculations, we model and relate the excitations observed, and pinpoint the role played by the environment of NiNc.

The measurements were performed in an ultra-high vacuum STM operating at $2.4$~K. The Cu(100) surface was cleaned \textit{in vacuo} by sputter/anneal cycles, while a sputter-cleaned etched tungsten tip was employed for tunneling. The tip was further prepared by controlled tip-surface contacts to ensure a monoatomically sharp copper apex. After submonolayer deposition of Nc onto the cold ($<100$~K) Cu(100) surface, well-ordered molecular assemblies on the surface were found [Fig.~\ref{fig1}(a)], along with isolated Nc molecules [Fig.~\ref{fig1}(b)]. The ring-shaped pattern in the images is produced by a cyclopentadienyl (Cp hereafter) ring and indicates that nickelocene is adsorbed with its principal axis perpendicular to the surface~\cite{Bachellier2016}. In the molecular network, however, these ``vertically'' adsorbed molecules coexist with ``horizontally'' adsorbed molecules (principal axis parallel to the surface), as sketched in the inset of Fig.~\ref{fig1}(a). This T-shaped configuration is governed by van der Waals interactions~\cite{Ormaza2015} and results in two possible molecular configurations, known as paired [Fig.~\ref{fig1}(a)] and compact (not shown)~\cite{Bachellier2016}. Our experimental observations regarding the formation and properties of the NiNc complex showed no significant difference between the two, therefore in the following we will only focus on the paired configuration.


To build NiNc complexes, we deposited single-nickel atoms from a Ni wire source (99.99\% purity) onto the cold surface ($\approx10$~K) through an opening in the cryostat shields. After exposure to a small amount of Ni atoms ($0.05$~monolayers), a new molecular species is present in the network [Fig.~\ref{fig1}(a)]. The molecular complex is imaged as a ring with an apparent height of $5.8\pm0.1$~{\AA} relative to the underlying copper surface [Figs.~\ref{fig1}(b) and~\ref{fig1}(c)], while the the neighboring Nc molecules have an apparent height of only $4.1\pm0.1$~{\AA}. Similar to the previous experiments of cobalt deposited onto a ferrocene network~\cite{Ormaza2015}, the ring-like shape demonstrates that the atom is positioned beneath a Nc molecule. This assignment is confirmed by our spin-polarized DFT calculations [Figs.~\ref{fig1}(d)\textemdash\ref{fig1}(f); details of the calculation are given as Supplemental Sec.~I] showing a $3$~eV energy difference in favor of a Ni atom beneath Nc rather than on top.  The Ni atom, which for clarity we refer to as Ni adatom hereafter, is located $2.4$~{\AA} above the copper surface. The NiNc complex displays a lower symmetry with a $0.5$~{\AA} misalignment between the two Ni atoms [Fig.~\ref{fig1}(d)] and a tilt of the principal axis of nickelocene [Fig.~\ref{fig1}(e)]. This tilt is confirmed by close-up STM images [see inset of Fig.~\ref{fig1}(a) and line profile in Fig.~\ref{fig1}(c)], differentiating the present NiNc complex from those investigated numerically in previous studies, where the Ni adatom is centered on the ring~\cite{Zelong2010,Morari2012}. We show below that NiNc adopts this structure outside the molecular network (see Fig.~\ref{fig4}). 

\begin{figure}
  \includegraphics[width=0.58\textwidth,clip]{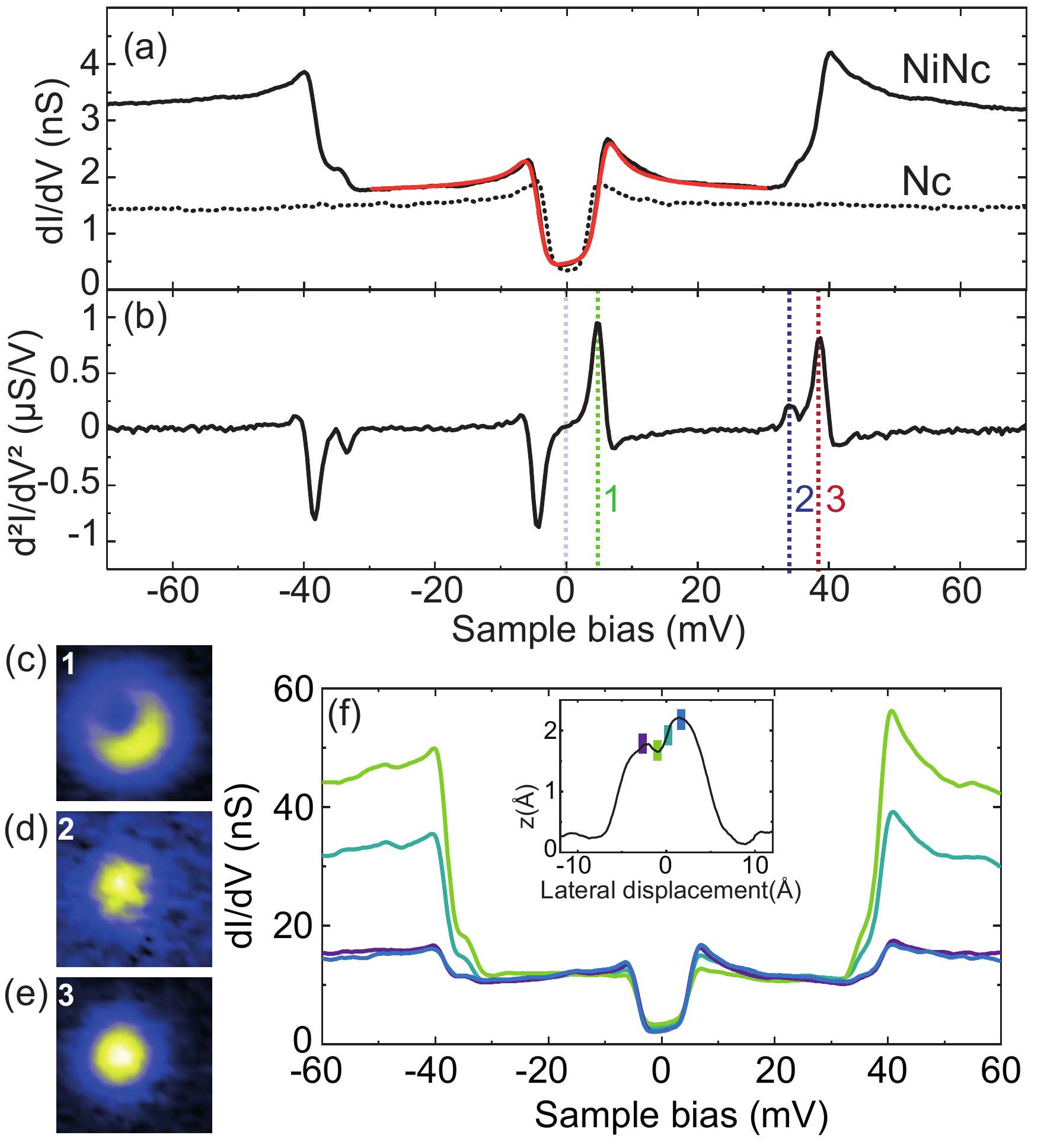}
  \caption{(a) $dI/dV$ spectrum and (b) $d^2I/dV^2$ spectrum acquired above the ring of a NiNc complex (feedback loop opened at $-80$~mV and $200$~pA). The solid red line is a fit based on a dynamical scattering model~\cite{Ternes2015}. The data exhibited a negligible tip dependence as reflected by the measurement uncertainties over a collection of NiNc complexes. The energy onsets are labeled $\mathbf{1}$, $\mathbf{2}$ and $\mathbf{3}$ in (b). The dashed curve in (a) is the $dI/dV$ spectrum acquired above the Cp ring of a Nc molecule (feedback loop opened at $-70$~mV and $100$~pA). Additional spectra of single Nc are presented in Fig.~S5. Panels (c), (d) and (e) present the spatial variation of the $d^2I/dV^2$ signal acquired at a constant-height and at a bias of $4.5$~mV, $32.5$~mV and $37$~mV, respectively (feedback loop opened at $-60$~mV and $500$ pA above a Nc molecule of the layer; 3 mV rms modulation amplitude). These biases correspond to the excitation thresholds highlighted in (b). The color scale in panels (c) and (e) correspond to a variation of $1.3$~$\mu$S/V, while in (d) the color scale corresponds to a variation of $0.4$~$\mu$S/V. (f) $dI/dV$ spectra acquired at four locations above NiNc [indicated by colored dots in the line profile of NiNc presented in the inset]. The feedback loop was opened at $-30$~mV ($300$~pA) above all locations in order to have same the amplitude for step $\mathbf{1}$ in the $dI/dV$ spectra. 
 \label{fig2}}
\end{figure}

Figure~\ref{fig2}(a) presents a typical $dI/dV$ spectrum acquired above the Cp ring of a NiNc complex in the paired network, while the $d^2I/dV^2$ spectrum is shown in Fig.~\ref{fig2}(b). All the spectra were recorded with a lock-in amplifier (200~$\mu$V rms and 716~Hz) using a tip that was verified to have a negligible electronic structure in the bias range investigated. The $dI/dV$ spectrum is dominated by stepped features, symmetric with respect to zero bias, which point to inelastic excitations. The energy onset of these steps, as determined over a collection of NiNc complexes, are $\lvert \epsilon_1\rvert=4.3\pm 0.4$~meV (the excitation is noted  $\mathbf{1}$ hereafter), $\lvert \epsilon_2\rvert=33.9\pm 0.5$~meV (noted $\mathbf{2}$) and $\lvert \epsilon_3\rvert=38.1\pm 0.6$~meV (noted $\mathbf{3}$). The data exhibited a negligible tip dependence. 

Given the similarity to the spin excitation spectrum measured above nickelocene [dashed line in Fig.~\ref{fig2}(a)], which is known from previous studies~\cite{Ormaza2017a,Ormaza2017b}, we assign $\mathbf{1}$ to a spin excitation. To confirm this assignment, we carried out spin-polarized DFT calculations (Supplemental Sec.~I, Fig.~S1). Using the relaxed structure of NiNc determined above, we find that the $d_{xz}$ and $d_{yz}$ frontier orbitals of Nc in the complex are spread out in a range of $\pm1$ eV around the Fermi level (Fig.~S1). The NiNc complex has a total magnetic moment of $1.4 $~$\mu_\text{B}$, corresponding to an antiferromagnetic coupling between the Ni adatom ($-0.2$~$\mu_\text{B}$) and Nc ($+1.6$~$\mu_\text{B}$) with a charge transfer of $0.1$ electrons from the substrate. The NiNc molecule has an effective spin of $1$. Consequently, we model $\mathbf{1}$ via a spin Hamiltonian that includes axial magnetic anisotropy
\begin{eqnarray}
\hat{H}_0=\hat{H}_A+DS_z^2,
\label{ham-1}
\end{eqnarray}
where $\hat{H}_A$ is an Anderson Hamiltonian involving a single nickelocene orbital of the NiNc complex (Supplemental Sec.~II). Within this viewpoint, $\mathbf{1}$ is assigned to a spin excitation occurring between the ground state $\ket{S=1, M=0}$ and the doubly degenerate $\ket{S=1, M=\pm1}$ excited states of NiNc [see Fig.~\ref{fig2}(a)], the onset $\lvert\epsilon_1\rvert$ corresponding to $D=4.6\pm0.2$~meV. The fit to the line shape based on Eq.~(\ref{ham-1}) is highly satisfactory [solid red line in Fig.~\ref{fig2}(a)]~\cite{Ternes2015}. The inclusion of many-body interactions in the fit through $\hat{H}_A$, \textit{i.e.} the inclusion of Kondo-like phenomena, is crucial for reproducing the cusp observed above the energy threshold of the spin excitation~\cite{Hurley2011,Korytar2012}. The cusp is associated to a Kondo fitting parameter $\mathcal{J}\rho=-0.4\pm0.2$ that is typical for nickelocene (Supplemental Sec.~III, Fig.~S2).


\begin{figure}
  \includegraphics[width=0.58\textwidth,clip]{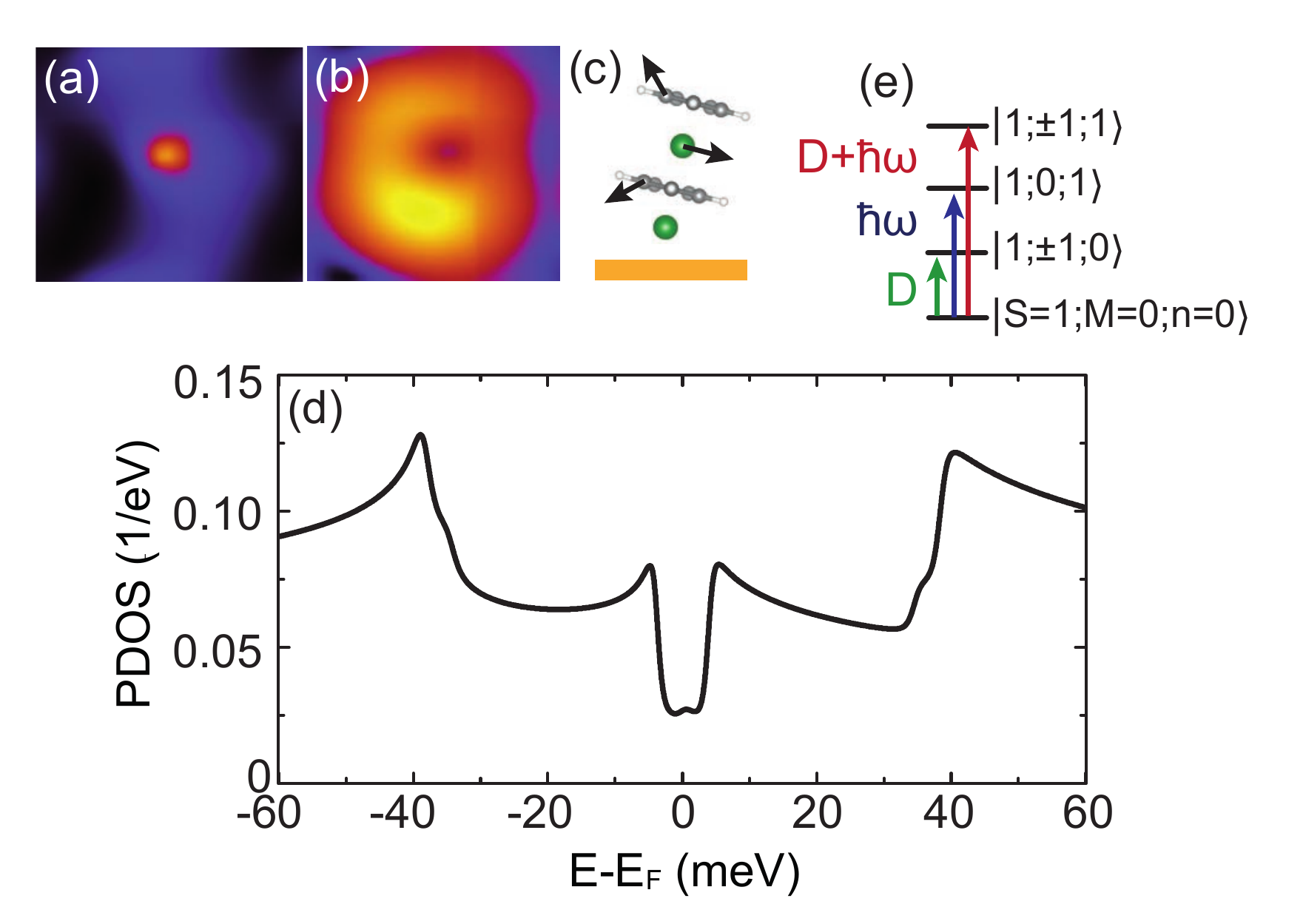}
  \caption{(a) Spatial dependence of the changes in conductance above $35$ mV in the presence of the electron-vibration coupling. The maximum change represents 5\% of the elastic conductance and is located in the center of the molecule. Black represents a zero change in conductance. (b) Calculated local density of states of a NiNc complex. (c) Sketch of the atomic motions involved in the vibrational mode. (d) Projected density of states on the occupied molecular orbital as a function of electron energy calculated from Eq.~(\ref{ham-2}). (e) State diagram of a NiNc complex based on Eq.~(\ref{ham-2}). The eigenstates are noted as $\ket{S,M; n}$ and the arrows depict the three transitions leading to the excitation steps in the $dI/dV$ spectrum.
\label{fig3}}
\end{figure}

Excitation $\mathbf{2}$ corresponds instead to a vibrational excitation of energy $\lvert \epsilon_2\rvert=\hbar\omega=33.9\pm 0.5$~meV. The spatial dependence of the $d^2I/dV^2$ signal above the NiNc molecule shows that while excitation $\mathbf{1}$ is located on the Cp ring [Fig.~\ref{fig2}(c)], excitation $\mathbf{2}$ is instead maximal in the center of the ring [Fig.~\ref{fig2}(d)]. To elucidate this difference, we computed the vibrational modes for the relaxed NiNc structure determined above. We use several schemes of finite difference (Supplemental Sec.~II) and consistently find that in a narrow energy window around the experimental energy $\lvert \epsilon_2\rvert$ there are three modes at $29.1$, $31.8$, and $35.5$~meV. The first mode corresponds to a molecular frustrated rotation about the Ni adatom, which we discard as at variance with the experimentally observed spatial location of the vibration. The second and third mode correspond to a translational motion of the Ni atom inside Nc. The tilted adsorption geometry of Nc on the Ni adatom breaks the degeneracy of these two modes observed in the gas phase. While the $31.8$ meV mode only gives a negligible change in the simulated conductance across the molecule, we find instead a dominant contribution for the $35.5$~meV mode with a spatial dependence matching experimental observations [Fig.~\ref{fig3}(a)]. The simulated STM image of NiNc is given in Fig.~\ref{fig3}(b) as a reference~\cite{Lorente2004}. Excitation $\mathbf{2}$ is then assigned to a Ni-Cp mode [Fig.~\ref{fig3}(c)] where the internal Ni atom moves parallel to the tilted Cp ring.

Excitation $\mathbf{3}$ shows a mixed behavior. First, all the spectra recorded so far showed that the energy of $\mathbf{3}$ is the sum of the energies of $\mathbf{1}$ and of $\mathbf{2}$, $\lvert \epsilon_3\rvert=\lvert \epsilon_1\rvert+\lvert \epsilon_2\rvert$ (Tab.~S1). Second, $\mathbf{3}$ bears spectroscopic signatures in common with both $\mathbf{1}$ and $\mathbf{2}$. On the one hand, $\mathbf{1}$ and $\mathbf{3}$ show similar amplitudes in the $dI/dV$ spectrum and exhibit a cusp above their corresponding excitation energies, which, as stressed above, is a characteristic feature of a spin excitation. On the other hand, $\mathbf{2}$ and $\mathbf{3}$ have the same spatial distribution over the NiNc complex [$dI^2/dV^2$ maps of Figs.~\ref{fig2}(d) and~\ref{fig2}(e)] and their step amplitudes in the $dI/dV$ spectra vary proportionally to one another across the NiNc molecule with a ratio of $3\pm1$ [Fig.~\ref{fig2}(f) and Supplemental Sec.~IV]. These results are remarkable and raise the question of how a second spin excitation can be present in the NiNc complex and how it relates to a vibrational excitation. 

In order to model the experimentally observed $dI/dV$ spectrum, we extend the spin Hamiltonian of Eq.~(\ref{ham-1}) to include vibrational effects
\begin{eqnarray}
\hat{H}=\hat{H}_0+\hbar \omega(\hat{b}^\dagger\hat{b}+\frac{1}{2})+W\sum_\sigma \hat{d}^{\sigma \dagger}\hat{d}^\sigma (\hat{b}+\hat{b}^\dagger). 
\label{ham-2}
\end{eqnarray}
As previously, we take a single $d$-level for the molecule and introduce the states $d$ annihilated and created by $\hat{d}^\sigma$ and $\hat{d}^{\sigma \dagger}$ with spin $\sigma$, respectively. We also assume that a molecular vibration of frequency $\omega$ is generated or annihilated by $\hat{b}^\dagger$ or $\hat{b}$ and, moreover, that the electron and vibration couple with a strength $W$ when the state $d$ is populated. A vibron is excited when an electron tunnels into the molecule. This in turn affects the electronic correlation included in $\hat{H}_A$, as well as the probability of spin excitation, which strongly depends on the occupation of the molecular orbital~\cite{Korytar2012}. The computed  $dI/dV$ spectrum is presented in Fig.~\ref{fig3}(d) (see Supplemental Sec.~II for details), where $D=4.6$ meV and $\hbar \omega=35.5$ meV. We use here a weak electron-vibration coupling of $W=20$ meV as estimated from the DFT calculations presented above~\cite{Monturet2008}. The results are robust to the parameters used for the molecular orbital. This model does not make use of any spin-vibron coupling. Only the electron-phonon interaction in the presence of spin excitations is able to explain the full spectra. We stress that the inclusion of dynamical electron correlations in the calculation is essential to correctly grasp the amplitude of the vibrational step. 

\begin{figure}
\includegraphics[width=0.58\textwidth,clip]{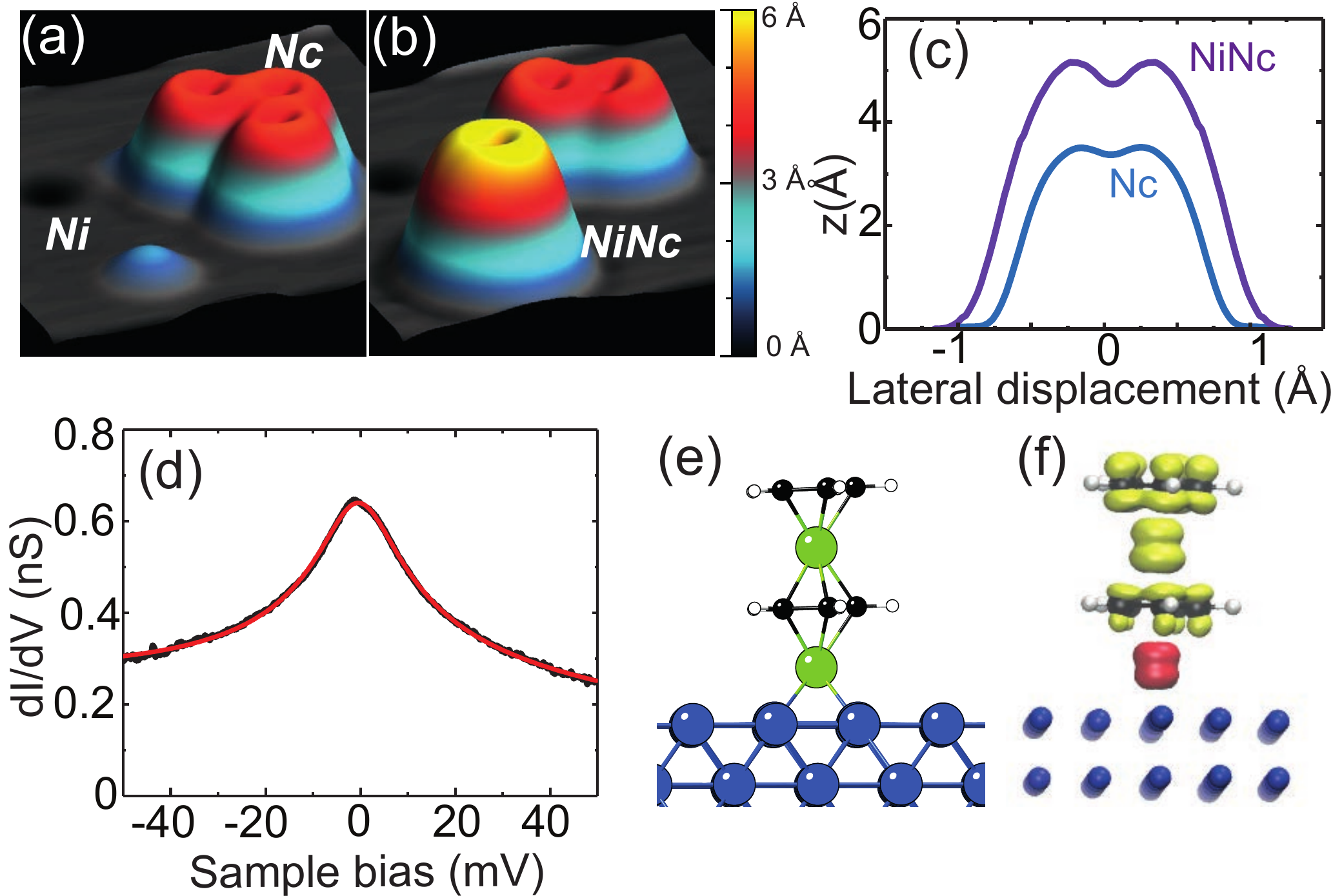}
  \caption{Panels (a) and (b) are pseudo 3D images showing the tip-assisted assembly of an isolated NiNc complex ($5\times 5$~nm$^2$, $-20$~mV, $20$~pA): (a) before and (b) after the transfer of Nc atop an isolated Ni adatom, and (c) height profiles of isolated NiNc and of isolated Nc. (c) $dI/dV$ spectrum measured above an isolated NiNc on Cu(100)  (feedback loop opened at $-50$~mV and $20$~nA). The solid red line is a Frota-Fano fit yielding a resonance centered at $-0.2\pm0.2$~meV. (e) DFT optimized structure of isolated NiNc on Cu(100), and (f) Side view of the spin density map.
\label{fig4}}
\end{figure}

Figure~\ref{fig3}(e) sketches the corresponding eigenenergies and allowed excitations. The first and second excited states correspond to a spin excitation $\ket{S=1,M=\pm1; n=0}$ and to a vibrational excitation $\ket{S=1,M=0; n=1}$, respectively. The third excited state, $\ket{S=1,M=\pm1; n=1}$, corresponds to a spin excitation energetically displaced upward in energy by a vibron. This excitation mechanism is similar to the coupled-spin vibrational Kondo effect~\cite{Paaske2005,Parks2007,Fernandez-Torrente2008,Rakhmilevitch2014}, where replicas of the Kondo resonance are found in the tunneling spectra at energies close to $\hbar\omega$. Consistent with this assignment, the relation between the three excitation thresholds simply reflects $\lvert \epsilon_3\rvert\simeq D+\hbar\omega$. The distortion of the NiNc molecule is negligible when the molecular vibration is active as expected for a weak electron-vibration coupling. A renormalized value of $D$ would be observed otherwise leading to $\lvert \epsilon_3\rvert\neq\lvert \epsilon_1\rvert+\lvert \epsilon_2\rvert$~\cite{Falk2011,Ruiz2012}. The relative amplitudes of the steps in the $dI/dV$ spectrum of NiNc can also be explained using this framework. Noting the step amplitudes by $\sigma_1$, $\sigma_2$ and $\sigma_3$ (Supplemental Sec.~IV, Fig.~S3), we find that all spectra recorded obey $\sigma_3/\sigma_0=(\sigma_2/\sigma_0)(\sigma_1/\sigma_0)$ (Tab.~S2), where the vacuum barrier thickness is accounted for by the elastic contribution $\sigma_0$. This relation indicates that the spin and the vibrational excitations occur independently one from another with transition rates proportional to $\sigma_1/\sigma_0$ and $\sigma_2/\sigma_0$, respectively, while the transition rate of the combined excitation $\sigma_3/\sigma_0$ is their product. This can lead, eventually, to a vibron-assisted spin excitation that exceeds in intensity the spin excitation [Fig.~\ref{fig2}(f)]. The same relation among step amplitudes was observed for single and double spin excitations produced by one electron tunneling across two magnetic molecules~\cite{Ormaza2017a}. 

To conclude, we highlight the importance of the nickelocene network for observing the inelastic excitations in the NiNc complex. For this purpose, we engineered the NiNc complex outside the network via a tip-assisted manipulation~\cite{Ormaza2016}. To do so, we first transferred an isolated Nc to the tip~\cite{Ormaza2017a,Ormaza2017b} and then transferred it back atop an isolated Ni adatom on the surface [Figs.~\ref{fig4}(a) and~\ref{fig4}(b)]. The newly formed molecule [Fig.~\ref{fig4}(b)] has an apparent height of $5.6\pm0.2$~{\AA} exceeding that of an isolated Nc molecule, $3.5\pm0.1$~{\AA} [Fig.~\ref{fig4}(c)], and presents a perfectly ring-shaped pattern, indicating that this complex is not tilted but lies straight. The $dI/dV$ spectrum changes completely compared to NiNc in the layer, showing now a broad resonance centered near the Fermi level [Fig.~\ref{fig4}(d)]; no inelastic excitation could be evidenced. DFT calculations give a clear picture of the new adsorption configuration adopted by NiNc. The Ni adatom, which is adsorbed in a hollow position of Cu(100), is now centered on the Cp ring [Fig.~\ref{fig4}(e)] and located at a distance of $2.47$~{\AA} from the copper surface. This higher adsorption symmetry for NiNc follows from the absence of steric constraints with neighboring nickelocenes. The lowest unoccupied molecular orbital of an isolated NiNc can be qualitatively represented by the mixing of the $p_z$ orbitals of C and of the $d_{xz}$ and $d_{yz}$ orbitals of Ni placed around the Fermi level. The calculated magnetic moment is $1.17~\mu_\text{B}$, which corresponds to an effective spin of $1/2$. The resonance of Fig.~\ref{fig4}(c) is then assigned to a spin-1/2 Kondo effect, a Frota-Fano fit~\cite{Pruser11} yielding a Kondo temperature of $T_\text{K}=68\pm7$~K. The Kondo effect is carried by the $d_{xz}$ and $d_{yz}$ frontier orbitals of Nc as 94\% of the spin density of NiNc is located on Nc. Compared to the NiNc complex of the network, the Ni adatom likely provides a hybridization pathway to the copper surface due to its central adsorption in the Cp ring, causing the effective spin of NiNc to drop, but concomitantly promoting the Kondo physics~\cite{Ormaza2017b}.

To summarize, we have shown that nickelocene adsorbed on a Ni atom yields a vibron-assisted spin excitation at an energy that is one order of magnitude higher than usual spin-excitation energies. This excitation is described through a model including axial magnetic anisotropy, intramolecular correlations and electron-phonon coupling. The vibron-assisted spin excitation should be observable in other molecular systems having magnetic anisotropy, although the vibrational excitation associated to it might not be easily detectable. This might be the case for some nickelocene molecules in the layer.

\begin{acknowledgments}
We thank M. Ternes for fruitful discussions and for providing his fitting program. This work was supported by the Agence Nationale de la Recherche (grants No. ANR-13-BS10-0016, ANR-11-LABX-0058 NIE and ANR-10-LABX-0026 CSC) and by the Agencia Espa\~{n}ola de Investigaci\'{o}n (grants No. MAT2016-78293-C6-1-R and MDM-2016-0618). M.-L.B. thanks the national computational center CINES and TGCC (GENCI project: A0030807364). 
\end{acknowledgments}

%

\end{document}